\documentclass[]{zakoproc}
\usepackage{graphicx}
\begin{document}

\title{Evolution of magnetic field in interacting galaxies}
\author{R. T. Drzazga, K. T. Chy\.zy, and W. Jurusik}
\institute{Astronomical Observatory, Jagiellonian University, ul. Orla 171, 30-244 Krak\'ow, Poland}
\markboth{Drzazga et al.}{Magnetic field evolution in interacting galaxies \ldots}

\maketitle

\begin{abstract}
Not much is currently known about how galaxy interactions affect an evolution 
of galactic magnetic fields. Here, for the first time, we explore a global 
evolution of magnetic fields with the advance of interaction process.
\end{abstract}

\section{Introduction}

Up to now, properties of magnetic fields were studied in detail only for one merging 
system -- the Antennae galaxies (Chy\.zy \& Beck~\cite{chyzy04}). This is one of 
the best known pair of interacting objects. Even though it is still debatable how 
advanced is this system in the process of merging. According to the most recent 
studies the galaxies are already after the second encounter and not after the first one, as 
was considered earlier (Karl et al.~\cite{karl10}). The most unusual magnetic fields 
in the Antennae system are those strong and regular ones in between galaxies -- in the 
overlapping region. Strong regular fields are also at the base of a tidal tail, from 
where they are stretched and transported along the tidal tail into the intergalactic medium. 
In this system magnetic fields are two times stronger than in typical spirals. 

The evolution of interacting galaxies was studied for the first time about 40 years ago by Toomre \& Toomre~(\cite{toomre72}). 
Their simulations of gravitational interactions between galaxies were used to construct a sequence of particular stages in the merging process, known now as a Toomre sequence (Toomre~\cite{toomre77}). It constitutes eleven pairs of galaxies arranged from earlier to later stages of interactions. For each pair a 
number from one to eleven is assigned. Numbers from one to nine  describe systems before the merger. 
The Antennae system is the first in this sequence. Number ten is assigned to a system at the stage 
of nuclear coalescence, and eleven describes the merger remnant.

\section{The sample}

The Toomre sequence was the base of our sample of interacting systems. In addition we selected galaxies 
from the sample constructed by Brassington et al.~(\cite{brassington07}) who investigated evolution 
of X-ray emission during the merging process. Furthermore, we performed some more general search for angularly 
large interacting galaxies and selected those for which radio data were available in the VLA archive.
In total we selected 16 interacting systems. For each of the selected objects we assigned a corresponding number 
from the Toomre sequence. We call it the Interaction Stage. To describe all 
our galaxies we extended the Toomre sequence to earlier stages of interactions and assigned them 
number $-1$ or 0. They are the weakest interacting objects which usually show only weak tidal 
distortions in optical light or in HI distribution. For galaxies at more advanced stages than 
the Toomre galaxies we assigned numbers 12 and 13. The last system with number 13 is a 
proto-elliptical galaxy, NGC\,1700 which was studied by Brassington et al.~(\cite{brassington07}).

\begin{figure}[t]
\begin{center}
 \includegraphics[width=.410\textwidth, clip = True]{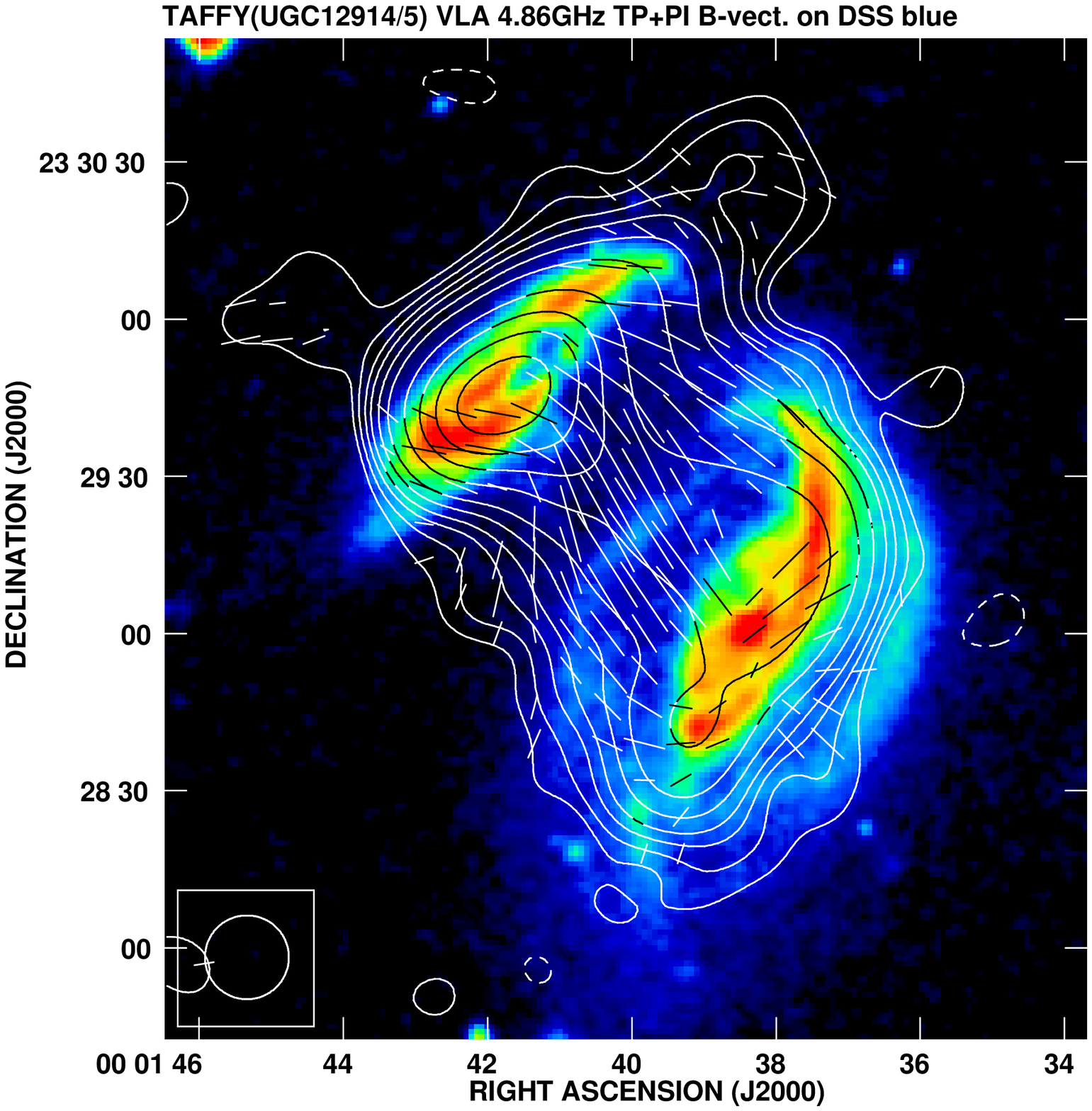}
 \includegraphics[width=.495\textwidth, clip = True]{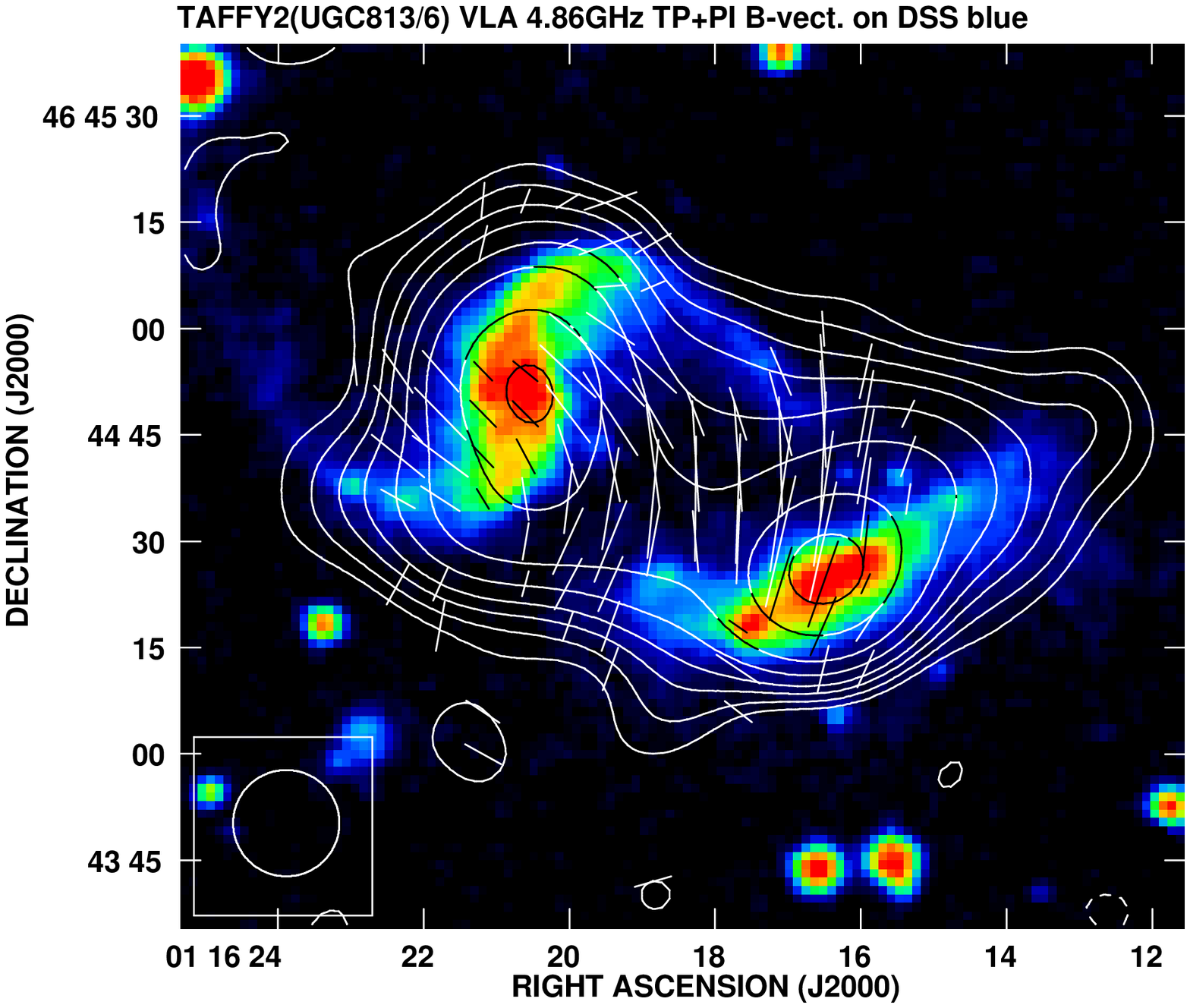}
 \caption{Left: Total power radio continuum map of the Taffy (UGC$\,$12914/UGC$\,$12915) at 4.86$\,$GHz with superimposed 
 B-vectors of polarization intensity, overlaid upon the DSS blue image. Right: Total power radio continuum map of the 
 Taffy2 (UGC$\,$813/UGC$\,$816) at 4.86$\,$GHz with superimposed B-vectors of polarization intensity, overlaid upon the DSS blue image.  
}
\end{center}
\label{TAFFY}
\end{figure}

\section{The Taffy and The Taffy2}

Ones of the most interesting interacting systems in our sample are Taffy and the Taffy2 galaxy pairs.
The unusual radio bridges between galaxies in these objects were discovered by Condon et al.~(\cite{condon93}) and Condon et al.~(\cite{condon02}). 
We re-reduced once again their radio observations to look for remnants of magnetic 
field structures in the galactic disks. 
We indeed found them (as in the Antennae system), they are especially well visible, in UGC\,12915 and similarly in UGC\,816 (Fig.~1). 
However, the most intriguing are the directions of B-vectors in 
both radio bridges of the taffies. The directions of B-vectors in the Taffy are aligned along the radio bridge, whereas 
in the Taffy2 they are perpendicular to the line joining both galaxies. 
There are three reasonable possibilities which can explain this finding: 
differences in ages of both systems, a difference 
in energy involved in collisions, and a complicated morphology of magnetic fields in 
the bridges. However, to evaluate these 
possibilities and solve that puzzle more detailed MHD simulations are needed.   

\section{Magnetic field strength and regularity}

The derived for our sample of interacting galaxies total magnetic field strengths are ranging from 5 $\mu$G for Arp\,222 to 27 $\mu$G for Arp\,220. These 
calculations were performed assuming the equipartition between magnetic fields and cosmic rays (Beck \& Krause~\cite{beck05}). 
The mean value of magnetic field computed for the whole sample is $14\pm5\,\mu$G. 
It is higher than the value reported by Niklas~(\cite{niklas95}) for his sample of galaxies of various Hubble types.
In contrast to the total magnetic field the strength of the regular field is like in other galaxies. 
However, the field regularity, which is the ratio of regular and random field components, is lower than in 
typical galaxies (cf. Beck et al.~\cite{beck96}). This implicates, that the random component 
of the magnetic field is more efficiently produced in interacting galaxies than in normal spirals. 
Or the regular component is destroyed more efficiently by processes related to strong star formation.

\begin{figure}[ht]
\begin{center}
 \includegraphics[width=.4525\textwidth, clip = True]{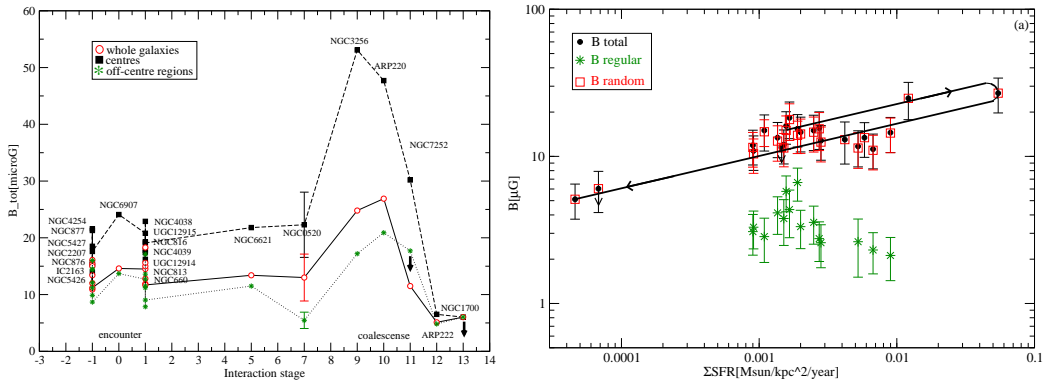}
 \includegraphics[width=.495\textwidth, clip = True]{Bsfr.eps}
 \caption{Left: The evolution of magnetic field strength in interacting galaxies. 
Mean magnetic field strengths are given for the whole galaxies (circles), 
centres (squares), and off-centre regions (asterisks). Arrows indicate upper limits of 
the field strength. Two special stages of interaction are denoted: the first 
galaxy encounter and the coalescence of the merger nuclei. The error bar for NGC\,520 is shown.
Right: Total magnetic field strengths (circles) versus surface density of SFRs for the 
whole interacting galaxies. The random and regular components of magnetic fields are also provided 
(squares and asterisks, respectively). An expected evolutionary track of interacting systems is 
also indicated (from Drzazga et al.~\cite{drzazga11}).}
\end{center}
\label{B_EVOL}
 \end{figure}

\section{Magnetic field evolution}

Are the strengths of galactic magnetic fields in any way related to the Interaction Stages?
In figure~2 the strength of total magnetic field obtained 
for our sample is plotted against the Interaction Stage. A general evolution of 
the magnetic fields is seen: for weakly interacting galaxies, 
shown in the left part of the plot, the magnetic field is nearly constant, with some dispersion. 
However, when interaction advances the magnetic field strength is increasing and 
reaches the maximum value for the systems being close to the stage of nuclear coalescence. 
After the nuclear coalescence the opposite trend is visible, the magnetic field is quickly weakening, indicating that the main processes responsible for generation of 
magnetic fields terminate at this stage. Such evolutionary trend is observable for 
the whole galaxies, their centres, and for off-centre regions. 
The largest enhancement of the magnetic field strengths is visible at galactic centres. 
This is in agreement with the finding of Hummel~(\cite{hummel81}) that the 
main increase of the radio emission of interacting galaxies, when compared to 
non-interacting objects, occurs in their centres. 

Similar evolutionary trend was also found by Georgakakis et al.~(\cite{georgakakis00}) for star formation efficiencies. 
This implicates, that the main source which regulates the generation processes of 
magnetic fields in the gravitationally interacting galaxies is related to the intensity of the star formation process. To check this hypothesis we determined relations between the total magnetic field 
strength, its random and regular components, and the star formation surface density 
($\Sigma SFR$) -- computed from the IRAS fluxes at 60 and 100 $\mu$m (Fig.~2).
Such dependencies were constructed for the whole galaxies, nuclear regions and off-centre regions (see, Drzazga et al.~\cite{drzazga11}). 
We found, that there is some weak relation for the total and random fields. 
The correlation coefficient determined for the whole disks and off-centre regions is about 0.49.
It is worth to note, that the obtained relations are strongly controlled by galaxies of the highest and lowest 
$\Sigma SFR$s. In the intermediate range of $\Sigma SFR$ galaxies with completely different interaction stages meet together, as indicated by 
the evolutionary track in figure~2. Thus, the full evolution of the magnetic field in the interacting 
galaxies can only be seen when they are arranged along the interaction stage. 

\section{Conclusions}

For the studied sample of 16 systems of gravitationally interacting galaxies 
we obtained the following results:

\begin{itemize}
\item[$\bullet$]{
the estimated mean of total magnetic field strength is $14\pm5\,\mu$G, which is larger than for the non-interacting objects -- this seems to be caused by enhanced production of random field component;
}
\item[$\bullet$]{
for the first time, we show a global evolution of magnetic fields with the
advance of interaction process. The main production of magnetic fields
terminates close after the nuclear coalescence. The strength of
magnetic fields in interacting galaxies is controlled by the intensity of 
star formation.
}
\end{itemize}

\acknowledgements{This work was supported by the Polish Ministry of Science and Higher
Education, grant 3033/B/H03/2008/35.}

\end{document}